\documentclass[12pt]{article}
\usepackage{amssymb} 
\usepackage{amsmath}

\begin{document}


\title{Applications of Lie systems in Quantum \\
Mechanics and Control Theory}

\author{Jos{\'e}\ F.\ Cari{\~n}ena\ and\ Arturo\ Ramos\\
Departamento de F\'{\i}sica Te\'orica,
Facultad de Ciencias\\
Universidad de Zaragoza, 50009, Zaragoza, Spain\\
E-mail: jfc@posta.unizar.es and arrg@wigner.unizar.es}
\date{}
\maketitle

\abstract{Some simple examples 
from quantum physics and control theory
are used to  illustrate  the application of
the theory of Lie systems. 

We will show, in particular, that for certain physical models
both of the corresponding classical and quantum problems
can be treated in a similar way, may be up to the replacement
of the involved Lie group by a central extension of it. 

The geometric techniques developed for dealing with 
Lie systems are also used in problems of control theory. 
Specifically, we will study some examples of control systems 
on Lie groups and homogeneous spaces.
}

\def\be{\begin{equation}}
\def\ee{\end{equation}}
\def\Eq#1{{\begin{equation} #1 \end{equation}}}
\def\R{{\Bbb R}}
\def\C{{\Bbb C}}
\def\Z{{\Bbb Z}}
\def\d{\partial}

\def\la#1{\lambda_{#1}}
\def\teet#1#2{\theta [\eta _{#1}] (#2)}
\def\tede#1{\theta [\delta](#1)}
\def\N{{\frak N}}
\def\Wei{\wp}

\font\frak=eufm10 scaled\magstep1

\def\ima{\hbox{{\rm Im}}}  
\def\bra#1{\langle#1|}
\def\ket#1{|#1\rangle}
\def\goth #1{\hbox{{\frak #1}}}
\def\<#1>{\langle#1\rangle}
\def\cotg{\mathop{\rm cotg}\nolimits}
\def\wt{\widetilde}
\def\const{\hbox{const}}
\def\grad{\mathop{\rm grad}\nolimits}
\def\Div{\mathop{\rm div}\nolimits}
\def\braket#1#2{\langle#1|#2\rangle}
\def\Erf{\mathop{\rm Erf}\nolimits}
\def\bq{{\bf q}}
\def\bp{{\bf p}}
\def\bQ{{\bf Q}}
\def\bP{{\bf P}}
\def\bx{{\bf x}}
\def\bby{{\bf y}}
\def\bz{{\bf z}}
\def\pd#1#2{\frac{\partial#1}{\partial#2}}
\def\smoy{\ast_{\! M}}
\def\Pol{\mathop{\rm Pol}\nolimits}
\def\pois#1#2{\{#1,#2\}}
\def\FL{{\cal F}L}
\def\pr{\pi}
\def\X{{\goth X}}
\def\Diff{{\hbox{Diff\,}}}

\def\GR{{\cal G}}
\def\matriz#1#2{\left( \begin{array}{#1} #2 \end{array}\right)}
\def\Map{\mathop{\rm Map}\nolimits}
\def\rank{\mathop{\rm rank}\nolimits}
\def\ad{\mathop{\rm ad}\nolimits}
\def\Ad{\mathop{\rm Ad}\nolimits}

\def\ba{\begin{eqnarray}}
\def\ea{\end{eqnarray}}
\def\bea{\begin{eqnarray}}
\def\eea{\end{eqnarray}}
\def\be{\begin{equation}}
\def\ee{\end{equation}}

\newtheorem{theorem}{Theorem}
\newtheorem{corollary}{Corollary}
\newtheorem{proposition}{Proposition}
\newtheorem{definition}{Definition}
\newtheorem{lemma}{Lemma}

\section*{1. Introduction: Lie systems.}

There exists a class of systems of time-dependent 
first order differential equations
\begin{equation}
\frac {dx^i}{dt}=X^i(x^1,\ldots,x^n,t),\quad i=1\ldots,n,  \label{dynsyst} 
\end{equation}
for which there is a function $\Phi:{\R}^{n(m+1)}\to {\R}^n$ such 
that the general solution can be written as
$x=\Phi(x_{(1)}, \ldots,x_{(m)};k_1,\ldots,k_n)$, 
where $\{x_{(j)}\mid j=1,\ldots,m\}$ is any set of particular 
but independent solutions of (\ref{dynsyst}), and $k_1,\ldots,k_n,$ 
are $n$ constants characterizing each particular solution.
These systems, to be called Lie systems,  
have been characterized by Lie \cite{LS} and are 
receiving much attention in recent years, both in 
physics and in mathematics \cite{CGM}-\cite{CarNas}. 
{}From the geometric viewpoint, Lie systems correspond
to $t$-dependent vector fields which can be 
written as a linear combination, with 
$t$-dependent coefficients, of a finite set of true vector 
fields closing on a finite-dimensional real 
Lie algebra \cite{CGM,CarRamGra}.

A simple instance of Lie system is the linear system
\begin{equation}
\frac {dx^i}{dt}=\sum_{j=1}^n A^i\ _j(t)\, x^j\ , \quad i=1,\ldots, n,\label{homsyst}
\end{equation}
for which the general solution can be written as a linear 
combination of $n$ independent particular solutions $x_{(1)}, \ldots ,x_{(n)}$,
i.e.,
$x=\Phi(x_{(1)},\ldots,x_{(n)},k_1,\ldots,k_n)
=k_{1}\, x_{(1)}+\cdots +k_{n}\,x_{(n)}$,
and in a similar way, the general solution for 
an inhomogeneous linear system can be written
as an affine function of $n+1$ independent 
particular solutions.

Another very remarkable example is the Riccati equation
$$
\frac{dx(t)}{dt}=a_2(t)\,x^2(t)+a_1(t)\,x(t)+ a_0(t)\ ,
$$
for which the superposition formula comes from the fact
that the cross ratio of four different solutions is a constant, 
see, e.g., \cite{CarRam}. 

The main point is that Lie systems are always related with 
Lie systems on Lie groups defined by right-invariant vector fields.
Let $G$ be a Lie group. If $\{a_1,\ldots,a_r\}$ is a basis of 
the tangent space $T_eG$ at the neutral element $e\in G$ 
and $X^R_\alpha$ denotes the right-invariant vector field in $G$ such that 
$X^R_\alpha(e)=a_\alpha$, a Lie system on $G$ will be written as
\begin{equation}
\dot g(t)=-\sum_{\alpha=1}^rb_\alpha(t)X^R_\alpha(g(t))\ .
\label{LiesinG}
\end{equation}
When applying $(R_{g(t)^{-1}})_{*g(t)}$ to both sides we obtain 
\begin{equation}
(R_{g(t)^{-1}})_{*g(t)}(\dot g(t))
=-\sum_{\alpha=1}^rb_\alpha(t)a_\alpha\ .  
\label{eqingr}
\end{equation}
which is usually written, with a slight abuse of notation, as
$$
(\dot g\, g^{-1})(t) =-\sum_{\alpha=1}^rb_\alpha(t)a_\alpha\ .
$$
This equation is right-invariant: if $\bar g(t)$ is a solution 
with initial condition $\bar g(0)=e$, the solution with initial 
condition $g(0)=g_0$ is given by $\bar g(t)g_0$. Therefore, we only 
need to find the solution of (\ref{eqingr}) starting from 
the neutral element.

Let $H$ be any closed subgroup of $G$, $M=G/H$ the corresponding 
homogeneous space, $\tau:G\to G/H$ the natural projection, 
and $\Phi:G\times M\to M$ the usual left action of $G$ on $M$.  
The right-invariant vector fields $X^R_\alpha$ are $\tau$-projectable 
onto the corresponding fundamental vector fields $-X_\alpha=-X_{a_\alpha}$, i.e., 
$$
\tau_{*g}X_\alpha^R(g)=-X_\alpha(gH)\ ,
$$
and therefore we have a Lie system on $M=G/H$ associated to (\ref{LiesinG}):
$$
\dot x(t)=\sum_{\alpha=1}^r b_\alpha(t)X_\alpha(x(t))\ ,
$$
where we denote $x=gH\in M$. Then, the solution of this system starting 
from $x_0$ is given by $x(t)=\Phi(g(t),x_0)$, where $g(t)$ is the 
solution of (\ref{LiesinG}) starting from the identity.
In this sense the equation (\ref{LiesinG}) has a universal character, 
and it will have an associated Lie system on each homogeneous space of $G$
\cite{CarRamGra}.

Lie systems are of interest not only in the theory of 
differential equations but also in other related fields.
For example, they are important in classical or even in quantum physics 
(for instance, in order to study the non-relativistic dynamics 
of a spin $1/2$ particle, when only the spinorial part 
is considered \cite{CGM,CGM01}). Another field where Lie systems
play an important r\^ole is in geometric control theory.

The aim of this article is to illustrate these applications.
Thus, after a brief account of a generalization of the method proposed by 
Wei and Norman \cite{CGM01,CarMarNas,CarRamcinc,WN1,WN2}, to be used later, 
we will study the particular case where the Lie systems of interest 
are Hamiltonian systems as well, both in the classical and quantum frameworks. 
The theory is illustrated through the particularly interesting example  
of generic classical and quantum quadratic time-dependent Hamiltonians.
In particular, we show that there exist $t$-dependent quantum systems
for which one is able to write in an explicit way the time evolution 
of any state of the system.
The very simple case of both the classical and quantum time-dependent 
linear potential will be explicitly solved. 
We will show as well the use of the theory of Lie systems in geometric
control theory, when dealing with drift-free systems that are linear 
in the control functions. In particular, we will study from this new perspective 
several well-known control systems: the robot unicycle \cite{NijSch90}, 
the Brockett nonholonomic integrator \cite{Bro82} and its realization in 
the model of a hopping robot in flight phase \cite{MurSas93},
and the kinematic equations of a generalization 
due to Jurdjevic \cite{Jur95} of the elastic problem of Euler.  

\section*{2. The  Wei and Norman method.}
Let $G$ be a Lie group as in the previous section.
We are interested in finding the curve $g(t)\in G$ such that
\begin{equation}
\dot g(t)\, g(t)^{-1}=-\sum_{\alpha=1}^r b_\alpha(t) a_\alpha\ ,
\label{eq_id_group}
\end{equation}
with $g(0)=e\in G$. We can use a method which is 
a generalization of the method proposed by Wei and Norman, 
in order to find the time-evolution operator for linear 
systems of type
$$
\frac {dU(t)}{dt}=H(t)U(t)\ ,
$$ 
with $U(0)=I$.
The generalized Wei--Norman method consists of writing the 
previous $g(t)$ in terms of a set of second kind canonical coordinates, 
$$
g(t)=\prod_{\alpha=1}^{r}
\exp(-v_\alpha(t)a_\alpha)=\exp(-v_1(t)a_1)\cdots\exp(-v_r(t)a_r)\ ,
$$
and transforming the equation (\ref{eq_id_group}) 
into a system of differential equations for the $v_\alpha(t)$,
with initial conditions $v_\alpha(0)=0$, $\alpha=1,\,\dots,\,r$. 
Such a system is obtained from the following relation:
\begin{equation}
\sum_{\alpha=1}^r \dot v_\alpha \left(\prod_{\beta<\alpha} 
\exp(-v_\beta(t) \ad(a_\beta))\right)a_\alpha
=\sum_{\alpha=1}^r b_\alpha(t) a_\alpha\,.
\label{eq_Wei_Nor}
\end{equation}
If the Lie algebra of $G$ is solvable, the solution of the 
previous system can be obtained by quadratures. If instead, the Lie 
algebra of $G$ is semi-simple, then the integrability by quadratures is not assured \cite{CGM01,CarMarNas,CarRamcinc,WN1,WN2}.

\section*{3. Hamiltonian systems of Lie type.}

An interesting and important case 
occurs when $(M,\Omega)$ is a symplectic 
manifold and the vector fields in $M$ arising in the expression 
of the $t$-dependent vector field describing a Lie system are Hamiltonian 
vector fields closing on a finite-dimensional real Lie algebra $\goth g$. 
These vector fields correspond to a symplectic action of a 
Lie group $G$ with Lie algebra $\goth g$ on the symplectic 
manifold $(M,\Omega)$. 

The Hamiltonian functions $h_\alpha$ of such vector fields,
defined by $i(X_\alpha)\Omega=-dh_\alpha$, in general 
do not close on the same Lie algebra $\goth g$ when the Poisson 
bracket is considered, since we can only assure that 
$$
d\left(\{h_\alpha,h_\beta\}-h_{[\alpha,\beta]}\right)=0\ ,
$$
and therefore, they span a Lie algebra extension of the original one.

The situation in quantum mechanics is quite similar:
the Hilbert space $\cal H$ can be seen as a real manifold with a global chart. 
The tangent space $T_\phi\cal H$ at any point $\phi\in\cal H$ 
can be identified with $\cal H$ itself, where the 
isomorphism which associates $\psi\in\cal H$ with the 
vector $\dot{\psi}\in T_\phi\cal H$ is given by: 
$$ 
\dot{\psi}f(\phi):= \left(\frac d{dt}f(\phi+t\psi)\right)_{|t=0}\ ,
\qquad\forall f\in C^\infty(\cal H)\ .
$$

The Hilbert space $\cal H$ is endowed with a symplectic 2-form $\Omega$
defined by 
$$ 
\Omega_{\phi}(\dot{\psi},\dot{\psi'})=2\,\ima\langle \psi|\psi' \rangle\ .
$$
A vector field is just a map $A\colon \cal H\to\cal H$; 
therefore a linear operator $A$ on $\cal H$ is a special kind of vector field.
Given a smooth function $a\colon\cal H\to\R$, its differential
$da_\phi $ at $\phi\in \cal H$ is an element of the (real) dual ${\cal H}'$ 
given by
$$
\langle da_\phi,\psi\rangle:=\left({\frac{d}{dt}}a(\phi+t\psi)\right)_{|t=0}\ .
$$
Actually, the skew-self-adjoint linear operators $-i\,A$ in $\cal H$,
for a self-adjoint operator $A$, define Hamiltonian vector fields, 
the Hamiltonian function of $-i\, A$\ being $a(\phi)=\frac 12 \<\phi,A\phi>$. 
Therefore, the Schr\"odinger equation plays the r\^ole of 
Hamilton equations, because it determines the integral curves 
of the vector field $-i\,H$, where $H$ is the Hamiltonian of the system
 \cite{BCG}.

In particular, the theory of Lie systems applies 
in the previous framework when we have a $t$-dependent 
quantum Hamiltonian that can be written as a linear combination, 
with $t$-dependent coefficients, of Hamiltonians $H_i$ closing on
a finite-dimensional real Lie algebra under the commutator bracket. 
However, note that this Lie algebra does not necessarily 
coincide with that of the corresponding classical problem, 
but it may be a Lie algebra extension of the latter.  

\section*{4. Time-dependent quadratic Hamiltonians.}

For the illustration of the classical and quantum situations described 
in the previous section, we consider now the important examples provided
by the time-dependent classical and quantum quadratic Hamiltonians.
 
The first one is the mechanical system for which the configuration 
space is the real line $\R$, 
the corresponding phase space $T^*\R$, endowed with its 
canonical symplectic structure $\omega=dq\wedge dp$, and 
the time-dependent classical Hamiltonian 
\begin{equation}
H=\alpha(t)\,\frac{p^2}2+\beta(t)\,\frac{q\,p}2+\gamma(t)\,\frac{q^2}2
+\delta(t)\,p+\epsilon(t)\,q\ .
\label{cgqH}
\end{equation}

The dynamical vector field solution of the dynamical equation
$$
i(\Gamma_H)\,\omega=dH\ ,
$$ 
is given by 
\begin{equation}
\Gamma_H=\left(\alpha(t)\,p+\frac 12 \beta(t)\,q+\delta(t)\right)\pd{}{q}
-\left(\frac 12 \beta(t)\,p+\gamma(t)\,q+\epsilon(t)\right)\pd{}{p}\ ,
\label{Ham_cuad_clas_vf}
\end{equation}
which can be rewritten as 
$$
\Gamma_H=\alpha(t)\,X_1+\beta(t)\,X_2+\gamma(t)\,X_3
-\delta(t)\,X_4+\epsilon(t)\,X_5\ ,
$$
with 
$$
X_1=p\,\pd{}q\,,\ X_2=\frac 12\left(q\,\pd {}q-p\,\pd{}p\right)\,,\
X_3=-q\,\pd{}p\,,$$
$$ X_4=-\pd{}q\,,\ X_5=-\pd{}p\,,
$$
being vector fields which satisfy the following commutation relations:
\begin{eqnarray}
&&[X_1,X_2]=X_1\,,\quad [X_1,X_3]=2\,X_2\,,\nonumber\\&& [X_1,X_4]=0\,,\qquad [X_1,X_5]=-X_4\,,\nonumber\\
&&[X_2,X_3]=X_3\,,\quad [X_2,X_4]=-\frac 12\,X_4\,,
\label{comm_rels_quadr_clas_hvf}                                                        \\
&& [X_2,X_5]=\frac 12\, X_5\,,\ [X_3,X_4]=X_5\,,\nonumber\\&& [X_3,X_5]=0\,,\qquad [X_4,X_5]=0\,,                      \nonumber
\end{eqnarray}
and therefore they close on a five-dimensional real Lie algebra. 
Consider the abstract, five-dimensional, Lie algebra $\goth g$ such that 
in a basis $\{a_1,\,a_2,\,a_3,\,a_4,\,a_5\}$, the Lie products are 
analogous to that of (\ref{comm_rels_quadr_clas_hvf}). 
Then, $\goth g$ is a semi-direct sum of the Abelian two-dimensional 
Lie algebra generated by $\{a_4,\,a_5\}$ with the 
${\goth{sl}}(2,\R)$ Lie algebra generated by $\{a_1,\,a_2,\,a_3\}$,
i.e., $\goth{g}=\R^{2}\rtimes{\goth{sl}}(2,\,\R)$. 
The corresponding Lie group will be the semi-direct product 
$G=T_2\odot SL(2,\R)$ relative to the linear action of $SL(2,\R)$ 
on the two-dimensional translation algebra.
When computing the flows of the previous vector fields $X_\alpha$, 
we see that they correspond to the affine action of $G$ on $\R^2$,
and therefore, the vector fields $X_\alpha$ can be regarded as 
fundamental fields with respect to that action, 
associated to the previous basis of the Lie algebra.
 
In order to find the time-evolution provided by the Hamiltonian (\ref{cgqH}),
i.e., the integral curves of the time-dependent vector 
field (\ref{Ham_cuad_clas_vf}), we can solve first the corresponding 
equation in the Lie group $G$ and then use the affine action of $G$ on $\R^2$.
We focus on the first of these questions: 
we should find the curve $g(t)$ in $G$ such that 
$$
\dot g\,g^{-1}=-\sum_{i=1}^5 b_i(t)\, a_i\ ,\qquad g(0)=e\,,
$$
with $b_1(t)=\alpha(t)$, $b_2(t)=\beta(t)$, 
$b_3(t)=\gamma(t)$, $b_4(t)=-\delta(t)$, and $b_5(t)=\epsilon(t)$.
The explicit calculation can be carried out by using the 
generalized Wei--Norman method, i.e., writing $g(t)$ in terms of a set of 
second class canonical coordinates, for instance, 
$$\begin{array}{ccl}
g(t)&=&\exp(-v_4(t)a_4)\exp(-v_5(t)a_5)\exp(-v_1(t)a_1)\cdot \\
&&\qquad\cdot\exp(-v_2(t)a_2)\exp(-v_3(t)a_3)\,,
\end{array}
$$
and then, a straightforward application of (\ref{eq_Wei_Nor}) leads to the system
\begin{eqnarray}
&&\dot v_1=b_1+b_2\, v_1+b_3\,v_1^2\ ,\quad             
\dot v_2=b_2+2\,b_3\,v_1\ ,\quad                        
\dot v_3=e^{v_2}\,b_3\ ,                        \nonumber\\
&&\dot v_4=b_4+\frac 12\, b_2\, v_4+b_1\,v_5\ ,
\quad\dot v_5=b_5-b_3\, v_4-\frac 12\, b_2\,v_5\ ,      \nonumber
\end{eqnarray}
with initial conditions $v_1(0)=\cdots=v_5(0)=0$.

For some specific choices of the functions $\alpha(t),\,\dots,\epsilon(t)$,
the problem becomes simpler and it may be enough to consider a subgroup, 
instead of the whole Lie group $G$, to deal with the arising system.
For instance, consider the classical Hamiltonian 
$$
H=\frac{p^2}{2 m}+f(t)\, q\ ,
$$
which in the notation of (\ref{cgqH}) has the only non-vanishing 
coefficients $\alpha(t)=1/m$ and $\epsilon(t)=f(t)$. 
Then, the problem is reduced to one in a three-dimensional subalgebra,
generated by $\{X_1,\,X_4,\,X_5\}$. The associated Lie group will be the
subgroup of $G$ generated by $\{a_1,a_4,a_5\}$. 
This example will be used later for illustrating the theory:
since such a subgroup is solvable, the problem can be 
integrated by quadratures.

Another remarkable property is that the Hamiltonian functions $h_\alpha$ 
corresponding to the Hamiltonian vector fields $X_1,\ldots, X_5$, 
defined by $i(X_\alpha)\omega=-dh_\alpha$, i.e.,
$$\begin{array}{ccl}
h_1(q,p)&=&-\frac {p^2}2\,,\quad h_2(q,p)=-\frac 12{q\,p}\,,\quad h_3(q,p)=-\frac {q^2}2\,,\\
 h_4(q,p)&=&p\,,
\qquad h_5(q,p)=-q\,,
\end{array}$$
have almost the same Poisson bracket relations as the vector fields $X_\alpha$,
but 
they  do not coincide
because of $\{h_4,h_5\}=1$, instead of $[X_4,X_5]=0$. In other words, they close on a Lie algebra which is a 
central extension of $\R^{2}\rtimes{\goth{sl}}(2,\,\R)$ 
by a one-dimensional algebra.

Let us now consider the quantum case \cite{KBW}, with applications 
in a number of physical problems, as for instance, the 
quantum motion of charged particles subject to time-dependent 
electromagnetic fields (see, e.g., \cite{FerMie94}), 
and connects with the theory of exact invariants developed by 
Lewis and  Riesenfeld (see \cite{LR69} and references therein). 

A generic time-dependent quadratic quantum Hamiltonian is given by 
\begin{equation}
H=\alpha(t)\,\frac{P^2}2+\beta(t)\,\frac{Q\,P+P\,Q}4+\gamma(t)\,\frac{Q^2}2+
\delta(t)P+\epsilon (t)\, Q+\phi(t) I\ .
\label{gqH}
\end{equation}
where $Q$ and $P$ are the position and momentum operators satisfying 
the commutation relation 
$$
[Q,P]=i\,I\ .
$$
The previous Hamiltonian can be written as a sum with $t$-dependent coefficients
$$
H=\alpha(t)\, H_1+\beta(t)\, H_2+\gamma(t)\,H_3
-\delta(t)\, H_4+\epsilon(t)\, H_5-\phi(t) H_6\ ,
$$
of the Hamiltonians
\begin{eqnarray}
H_1&=&\frac {P^2}2\,,\quad H_2= \frac 14 (QP+P\,Q)\,,\quad H_3=\frac {Q^2}2\,,\nonumber\\ 
H_4&=&-P\,,\quad H_5=Q\,,\quad H_6=-I\,,\nonumber
\end{eqnarray}
which satisfy the commutation relations
\begin{eqnarray}
&& [i H_1, i H_2]=i H_1\,,\quad [i H_1, i H_3]=2\,i H_2\,,\quad [i H_1, i H_5]=-i H_4\,,\nonumber\\
&& [i H_2, i H_3]=i H_3\,, \quad  [i H_2, i H_4]=-\frac i2\,H_4\,,\quad [i H_2, i H_5]=\frac i2\,H_5\,,\nonumber\\
&& [i H_3, i H_4]=i H_5\,,\quad [i H_4, i H_5]=i H_6\,,              \nonumber
\end{eqnarray}
and $$[i H_1, i H_4]=[i H_3,i H_5]=[i H_\alpha, i H_6]=0\,,\qquad  \alpha=1,\,\dots,\,5\ .$$ 
That is, the skew-self-adjoint operators $i H_\alpha$ generate 
a six-dimensional real Lie algebra which is a central extension of the 
Lie algebra arising in the classical case, $\R^{2}\rtimes{\goth{sl}}(2,\,\R)$, 
by a one-dimensional Lie algebra. It can be identified as the semi-direct sum of the 
Heisenberg--Weyl Lie algebra $\goth{h}(3)$, which is an ideal in the total Lie algebra, 
with the Lie subalgebra $\goth{sl}(2,\R)$, i.e., $\goth{h}(3)\rtimes{\goth{sl}}(2,\,\R)$.
Sometimes this Lie algebra is referred to as the extended symplectic 
Lie algebra $\goth{hsp}(2,\R)=\goth{h}(3)\rtimes{\goth{sp}}(2,\,\R)$.
The corresponding Lie group is the semi-direct product 
$H(3)\odot SL(2,\R)$ of the Heisenberg--Weyl group $H(3)$ with $SL(2,\R)$, 
see also \cite{KBW}.

The time-evolution of a quantum system 
can be described in terms of the evolution operator $U(t)$
which satisfies the Schr\"odinger equation (see, e.g., \cite{CohDiuLal77})
$$
i\frac{d U}{dt}=H(t) U\,,\qquad U(0)=\mbox{Id}\,,
$$
where $H(t)$ is the Hamiltonian of the system. 
In our current case, the Hamiltonian is given by (\ref{gqH}),
and therefore the time-evolution of the system is given 
by an equation of the type
\begin{equation}
\dot g\,g^{-1}=-\sum_{\alpha=1}^6 b_\alpha(t)\, a_\alpha\ ,\qquad g(0)=e\,,
\label{eq_WN_qqh}
\end{equation}
with the identification of $g(t)$ with $U(t)$, $e$ with $\mbox{Id}$, 
$i H_\alpha$ with $a_\alpha$ for $\alpha\in\{1,\,\dots,\,6\}$ and the 
time-dependent coefficients $b_\alpha(t)$ are given by  
\begin{eqnarray}
&&b_1(t)=\alpha(t)\,,\ \,\,b_2(t)=\beta(t)\,,\ b_3(t)=\gamma(t)\,,  \nonumber\\
&&b_4(t)=-\delta(t)\,,\ b_5(t)=\epsilon(t)\,,\ b_6(t)=-\phi(t)\,. \nonumber
\end{eqnarray}

We would like to remark that time-dependent quantum Hamiltonians  
are seldom studied, because it is generally difficult to find 
their time evolution. However, in the case the system could be 
treated as a Lie system in a certain Lie group, the calculation
of the evolution operator is reduced to the problem of 
integrating the system appearing after the application of the 
Wei--Norman method. In the case the associated Lie group is solvable,
the integration can be made by quadratures, leading to an exact 
solution of the problem. We will see an example in the next section.

The calculation of the solution of (\ref{eq_WN_qqh}) 
can be carried out by using the generalized Wei--Norman method, i.e., 
writing $g(t)$ in terms of a set of second class canonical coordinates.
We take, for instance, the factorization 
\begin{eqnarray}
&&g(t)=\exp(-v_4(t)a_4)\exp(-v_5(t)a_5)\exp(-v_6(t)a_6) \nonumber\\
&&\hskip10truemm\times\exp(-v_1(t)a_1)\exp(-v_2(t)a_2)\exp(-v_3(t)a_3)\ ,       \nonumber
\end{eqnarray}
and therefore, the equation (\ref{eq_Wei_Nor}) leads in this case to the
system 
\begin{eqnarray}
&&\dot v_1=b_1+b_2\, v_1+b_3\,v_1^2\ ,\quad             
\dot v_2=b_2+2\,b_3\,v_1\ ,\quad                        
\dot v_3=e^{v_2}\,b_3\ ,                        \nonumber\\
&&\dot v_4=b_4+\frac 12\, b_2\, v_4+b_1\,v_5\ ,
\quad\dot v_5=b_5-b_3\, v_4-\frac 12\, b_2\,v_5\ ,      \nonumber\\     
&&\dot v_6=b_6+b_5\, v_4-\frac 12\, b_3\,v_4^2+\frac 12\, b_1\,v_5^2\ ,\nonumber
\end{eqnarray}
with initial conditions $v_1(0)=\cdots=v_6(0)=0$.

Analogously to what happened in the classical case, special choices of the 
time-dependent coefficient functions $\alpha(t),\,\dots,\phi(t)$ may lead to
problems for which the associated Lie algebra is a subalgebra of that of the
complete system, and similarly for the Lie groups involved. For example,
we could consider as well the quantum Hamiltonian linear in the positions
$$
H=\frac{P^2}{2 m}+f(t)\, Q\ ,
$$
which in the notation of (\ref{gqH}) has the only non-vanishing 
coefficients $\alpha(t)=1/m$ and $\epsilon(t)=f(t)$. This problem
can be regarded as a Lie system associated to the four-dimensional 
Lie algebra generated by $\{iH_1,\,iH_4,\,iH_5,\,iH_6\}$, which 
is also solvable, and hence the problem can be solved by quadratures. 

The treatment of this system, as well as that of its classical version,
according to the theory of Lie systems, is the subject of the next section.

\section*{5. An example: classical and quantum time-dependent linear potential.}

Let us consider the classical system described by the classical Hamiltonian
\begin{equation}
H_c=\frac{p^2}{2m}+f(t)\, q\ ,
\label{Ham_td_lin_pot_cl}
\end{equation}
and the corresponding quantum Hamiltonian
\begin{equation}
H_q=\frac{P^2}{2m}+f(t)\, Q\ ,
\label{Ham_td_lin_pot_qu}
\end{equation}
describing, for instance when $f(t)=e\, E_0+e\,E\,\cos\omega t $, 
the motion of a particle of electric charge $e$ 
and mass $m$ driven by a monochromatic electric field. 

We will study in parallel the classical and the quantum problems
by reduction of both of them to similar equations, 
and solving them by the generalized Wei--Norman method.
The only difference between the two cases is that the
Lie algebra arising in the quantum problem is a central extension
of that of the classical one.

The classical Hamilton equations of motion 
for the Hamiltonian (\ref{Ham_td_lin_pot_cl}) are
\begin{equation}
\left\{\begin{array}{cl}{\dot q}&={\displaystyle{\frac pm}}\,,\cr
{\dot p}&=-f(t)\,,\end{array}\right.
\label{eqsHam_Ham_td_lin_pot_cl}
\end{equation}
and therefore, the motion is given by
\begin{eqnarray}
q(t)&=&q_0+\frac{p_0\, t}m-\frac 1m\int_0^tdt'\,\int_0^{t'}f(t'')\, dt''\ ,\cr
p(t)&=&p_0-\int_0^tf(t')\,dt'\ .                                
\label{emls}
\end{eqnarray}

The $t$-dependent vector field describing the time evolution,
$$
X=\frac pm\,\pd{}q-f(t)\,\pd{}p\ .
$$
can be written as a linear combination
$$
X=\frac 1m\, X_1-f(t)\, X_2\ ,
$$
with
$X_1=p\,\partial/\partial q$, and $X_2=\partial/\partial p$
being vector fields closing on a 3-dimensional Lie algebra with 
$X_3=\partial/\partial q$,
isomorphic to the Heisenberg--Weyl algebra, namely,
$$
[X_1,X_2]=-X_3\ , \qquad [X_1,X_3]=0\ ,\qquad [X_2,X_3]=0\ .
$$
The flow of these vector fields is given, respectively, by
\begin{eqnarray}
&&\phi_1(t,(q_0,p_0))=(q_0+p_0\, t,p_0)\,,      \nonumber\\
&&\phi_2(t,(q_0,p_0))=(q_0,p_0+t)\,,            \nonumber\\
&&\phi_3(t,(q_0,p_0))=(q_0+t,p_0)\,.            \nonumber
\end{eqnarray}
In other words, $\{X_1,\,X_2,\,X_3\}$ are fundamental vector fields
with respect to the action of the Heisenberg--Weyl 
group $H(3)$, realized as the Lie group of upper triangular $3\times 3$ matrices, 
on $\R^2$ given by
$$
\matriz{c}{\bar q\\ \bar p\\1}=\matriz{ccc}{1&a_1&a_3\\0&1&a_2\\0&0&1}
\matriz{c}{q\\ p\\1}\ .
$$

Note that  $X_1$, $X_2$ and $X_3$ are 
Hamiltonian vector fields with respect to the usual symplectic structure, 
$\Omega=dq\wedge dp$, meanwhile the corresponding Hamiltonian 
functions $h_\alpha$ such that $i(X_\alpha)\Omega=-dh_\alpha$ are
$$
h_1=-\frac {p^2}2\,,\qquad h_2=q\,,\qquad h_3=-p\ ,
$$
and therefore
$$
\{h_1,h_2\}=-h_3\,, \quad \{h_1,h_3\}=0\,, \quad \{h_2,h_3\}=-1\,. 
$$
Then, the functions $\{h_1,\,h_2,\,h_3\}$, jointly with $h_4=1$, close on
a four-dimensional Lie algebra under the Poisson bracket which is 
a central extension of that generated by $\{X_1,\,X_2,\,X_3\}$.

If $\{a_1,\,a_2,\,a_3\}$ is a basis of the Lie algebra with 
non-vanishing defining relations $[a_1,a_2]=-a_3$, 
the corresponding equation in the group $H(3)$ to 
the system (\ref{eqsHam_Ham_td_lin_pot_cl})
reads
$$
\dot g\, g^{-1}=-\frac 1 m\, a_1+f(t)\, a_2\ .
$$
Using the Wei--Norman formula (\ref{eq_Wei_Nor}) with 
$$g=\exp(-u_3\, a_3)\,\exp(-u_2\, a_2)\,\exp(-u_1\, a_1)$$
we arrive to the system of differential equations
$$
\dot u_1=\frac 1m\ ,\qquad \dot u_2=-f(t)\ ,\qquad
\dot u_3-\dot u_1\, u_2=0\ ,
$$
together with the initial conditions
$u_1(0)=u_2(0)=u_3(0)=0$,
with solution 
$$
u_1=\frac tm\ ,\quad u_2=-\int_0^tf(t')\, dt'\ ,
\quad u_3=-\frac 1m \int_0^t dt'\int_0^{t'}f(t'')\, dt''\ .
$$
Therefore, the motion is given by
$$
\matriz{c}{q\\p\\1}
=\matriz{ccc}{1&\frac tm&-\frac 1m\int_0^t dt'\int_0^{t'}f(t'')\,dt''
\\0&1&-\int_0^tf(t')\, dt'
\\0&0&1}
\matriz{c}{q_0\\p_0\\1}\ ,
$$
in agreement with (\ref{emls}).
We can immediately identify the constants of motion 
\begin{eqnarray}
I_1&=&p(t)+\int_0^tf(t')\, dt'\ ,\cr
I_2&=&q(t)-\frac 1m\left(p(t)+\int_0^tf(t')\, dt'\right) t
+\frac 1m\int_0^tdt'\int_0^{t'}f(t'')\, dt''\ .\nonumber
\end{eqnarray}

As far as the quantum problem is concerned, 
notice that the quantum Hamiltonian $H_q$ may be written as a sum
$$
H_q=\frac 1 m \, H_1-f(t)\, H_2\ ,
$$
with
$$
H_1=\frac {P^2}2\ ,\qquad  H_2=-Q\ .
$$

The skew-self-adjoint operators $-i\, H_1$ and $-i\, H_2$ close on a four-dimensional 
Lie algebra with $-i\, H_3=-i\,P$, and $-i H_4=i\,I$, 
isomorphic to the above mentioned central extension of 
the Heisenberg--Weyl Lie algebra,
$$
[-iH_1,-iH_2]\!=\!-\!iH_3\,,\ [-iH_1,-iH_3]\!=\!0\,,\ [-iH_2,-iH_3]\!=\!-\!iH_4\,.
$$

As we have seen in the preceding section, the time-evolution of 
our current system is described by means of the evolution operator $U$,
which satisfies 
$$
\frac{d U}{dt}=-i H_q U\,,\qquad U(0)=\mbox{Id}\,.
$$
This equation can be identified as that of a Lie system in a
Lie group such that its Lie algebra is the one mentioned above.
Let $\{a_1,\,a_2,\,a_3,\,a_4\}$ be a basis of the Lie algebra with 
non-vanishing defining relations $[a_1,a_2]=a_3$ and $[a_2,a_3]=a_4$.
The equation in the group to be considered now is
$$
\dot g\, g^{-1}=-\frac 1 m\, a_1+f(t)\, a_2\ .
$$
Using  
$g=\exp(-u_4\, a_4)\exp(-u_3\, a_3)\,\exp(-u_2\, a_2)\,\exp(-u_1\, a_1)$, 
the Wei--Norman method provides the following equations:
\begin{eqnarray}
&\dot u_1=\frac 1m\,, \quad\quad &\dot u_2=-f(t)\,,                     \nonumber\\
&\quad\quad\dot u_3=-\frac 1m\,u_2\,, \quad\quad &\dot u_4=f(t)\, u_3+\frac 1{2m}\, u_2^2\,,
                                                                \nonumber
\end{eqnarray}
together with the initial conditions 
$u_1(0)=u_2(0)=u_3(0)=u_4(0)=0$, whose solution is 
$$
u_1(t)=\frac tm\,,\quad u_2(t)=-\int_0^tf(t')\, dt'\,,
\quad u_3(t)=\frac 1m\int_0^tdt'\int_0^{t'}f(t'')\, dt''\,,
$$ 
and 
$$
u_4(t)=\frac 1m\!\int_0^t\! dt' f(t')\!\int_0^{t'}\!dt''\!\int_0^{t''}\!f(t''')\,dt'''
+\frac 1{2m}\!\int_0^t \!dt'\!\left(\!\int_0^{t'}\!dt''f(t'')\!\right)^2\ .
$$
These functions provide the explicit form of the time-evolution operator:
$$
\!U(t,\!0)\!=\!\exp(\!-i u_4(t))\!\exp(i u_3(t) P)\!\exp(\!-i u_2(t) Q)\!\exp(i u_1(t)\!P^2/2)\ .
$$

However, in order to find the expression of the wave-function in a
simple way, it is advantageous to use instead the factorization 
$$
g=\exp(-v_4\, a_4)\exp(-v_2\, a_2)\,\exp(-v_3\, a_3)\,\exp(-v_1\, a_1)\,.
$$
In such a case, the Wei--Norman method gives the system
\begin{eqnarray}
&\dot v_1=\frac 1m\,, \quad\quad &\dot v_2=-f(t)\,,                     \nonumber\\
&\quad\quad\dot v_3=-\frac 1m\,v_2\,, \quad\quad &\dot v_4=-\frac 1{2m} v_2^2\,,
                                                                \nonumber
\end{eqnarray}
jointly with the initial conditions $v_1(0)=v_2(0)=v_3(0)=v_4(0)=0$. The solution is
\begin{eqnarray}
&&v_1(t)=\frac tm\,,\quad v_2(t)=-\int_0^t dt'\,f(t')\,,\quad    \cr 
&&v_3(t)=\frac 1 m \int_0^t dt'\int_0^{t'} dt'' f(t'')\,,         \cr 
&&v_4(t)=-\frac 1{2m}\int_0^t dt'\left(\int_0^{t'} dt'' f(t'')\right)^2\,.
\nonumber
\end{eqnarray}
Then, applying the evolution operator on the initial 
wave-function $\phi(p,0)$, which is assumed to be written in momentum representation, we have
\begin{eqnarray}
\phi(p,t)&=&U(t,0)\phi(p,0)                                                     \nonumber\\
&=&\!\exp(\!-i v_4(t))\!\exp(\!-i v_2(t) Q)\!\exp(i v_3(t) P)\exp(i v_1(t)\!P^2/2)\!\phi(p,0) 
                                                                                \nonumber\\
&=&\exp(-i v_4(t))\exp(-i v_2(t) Q)e^{i(v_3(t) p+v_1(t) p^2/2)}\phi(p,0)        \nonumber\\
&=&\exp(-i v_4(t))e^{i(v_3(t) (p+v_2(t))+v_1(t) (p+v_2(t))^2/2)}\phi(p+v_2(t),0)\,, \nonumber
\end{eqnarray}
where the functions $v_i(t)$ are given by the preceding equations.

\section*{6. Applications in control theory.}

Control systems are described by systems of differential  equations
\begin{equation}
\frac{dx^i}{dt}=F(x^i,u^\alpha), \quad i=1,\ldots, n,\quad \alpha =1, \ldots,m,
\label{cont_syst}
\end{equation}
where $u^\alpha$ are the so-called control functions or simply controls, 
which are to be determined in such a way that, e.g., 
the trajectory passes through one or two specific points in the 
configuration space, or maybe gives some cost functional a
stationary value. 

A control system is said to be controllable if for any given initial
point $p$ there exists an integral curve of the corresponding vector 
field along $\pi:{\R}^{n+m}\to {\R}^{n}$ such that $(\pi\circ \gamma)(0)=p$, 
and a value $t_1$ of the parameter of the curve $\gamma$ 
such that $(\pi\circ \gamma)(t_1)=q$ for any final point $q$.

Consider the case of drift-free systems, 
linear in the control functions $u^\alpha(t)$, for which the
time-dependent vector field, whose integral curves are the solutions of (\ref{cont_syst}),
is
$$
X(t,x)=u^1(t)X_1(x)+\cdots+u^r(t)X_r(x)\,.
$$
Lie systems arise when the vector fields $X_\alpha$ 
close on a finite-dimensional real Lie algebra. 
The cases in which the $X_\alpha$ are either right-invariant vector 
fields in a certain Lie group $G$, or vector fields in a homogeneous
space of $G$, can be dealt with according to the theory of Lie systems:
reducing the problem to solve an equation on the Lie group $G$, of the form
$$
\dot g(t)=\sum_{\alpha=1}^r u_\alpha(t)X_\alpha(g(t))\ .
$$

Controllability of control systems on Lie groups 
has been analyzed by Brockett  \cite{Bro72} and  
Jurdjevic and Sussmann \cite{JurSus72}. 
It can be determined by studying algebraic properties of the 
corresponding Lie algebra $\goth g$. 

{\bf Theorem:} {\it
A drift-free right invariant system on a
connected Lie group $G$ is controllable if and only if the Lie algebra 
generated by  $\{X_1,\ldots, X_r\}$ is $\goth g$. 
}

{\sl Proof:} 
If $\goth h$ is the Lie algebra generated by $\{X_1,\ldots, X_r\}$,
then the Lie algebra of the Lie system is not $\goth g$ but the 
subalgebra $\goth h$. The orbit of the neutral element  
$e\in G$ is the subgroup $H$ of $G$ with Lie subalgebra $\goth h$. 
It is then clear that if $\goth h$ is a proper subalgebra of $G$, 
the system is not controllable, while it is so when ${\goth h}={\goth g}$.

\medskip
As examples of application of the theory of Lie systems 
to specific systems treated in control theory, we will study
several well-known systems: the robot unicycle \cite{LafSus91,NijSch90}, 
the Brockett system termed sometimes as Brockett nonholonomic 
integrator \cite{Bro82}, and a system which under certain 
approximation can be reduced to the former,
i.e., the model of a hopping robot in flight phase \cite{MurSas93}. 
Afterwards, we will study the kinematic equations of a generalization due 
to Jurdjevic \cite{Jur95} of the elastic problem of Euler, and finally we 
will briefly show how the reduction theory of Lie systems can be
 applied to two of these examples.    

\subsection*{{\bf 6.1.} Robot unicycle or model of an automobile as a Lie system.}

Our first example corresponds to the robot unicycle (see, e.g., \cite{LafSus91}).
Essentially, the same control system arises in a very simplified model of 
maneuvering an automobile \cite{NijSch90}. 

The configuration space is $\R^2\times S^1$, with coordinates $(x_1,\,x_2,\,x_3)$. 
The control system can be written as
\begin{equation}
\dot x_1=b_2(t)\sin x_3\,,\quad \dot x_2=b_2(t)\cos x_3\,,\quad \dot x_3=b_1(t)\,,
\label{cs_unicy}
\end{equation}
where $b_1(t)$ and $b_2(t)$ are the control functions. Its solutions are the 
integral curves of $b_1(t)\, X_1(x)+b_2(t)\, X_2(x)$, where
$$
X_1=\pd{}{x_3}\,,\quad\quad X_2=\sin x_3\,\pd{}{x_1}+\cos x_3\,\pd{}{x_2}\,.
$$
The Lie bracket of both vector fields, 
$$
X_3=[X_1,\,X_2]
=\cos x_3\,\pd{}{x_1}-\sin x_3\,\pd{}{x_2}\ ,
$$
is linearly independent from
$X_1,\,X_2$. They  satisfy 
$$
[X_1,\,X_2]=X_3\,, \quad [X_2,\,X_3]=0\,,\quad [X_1,\,X_3]=-X_2\,,
$$ 
therefore closing on a Lie algebra isomorphic to $\goth{se}(2)$. 
This Lie algebra has
a basis $\{a_1,\,a_2,\,a_3\}$ for which  
$$[a_1,\,a_2]=a_3\,, \quad [a_2,\,a_3]=0\,,\quad [a_1,\,a_3]=-a_2\ .$$

Writing the solution of the associated problem in  $SE(2)$ as
$$
g(t)=\exp(-v_1(t)a_1)\exp(-v_2(t)a_2)\exp(-v_3(t)a_3)\ ,
$$ 
then the Wei-Norman method leads to the system
$$
\dot v_1=b_1\,,\quad\dot v_2=b_2\,\cos v_1\,,\quad\dot v_3=b_2\,\sin v_1\,,
$$
with $v_1(0)=v_2(0)=v_3(0)=0$.
Denoting $B_1(t)=\int_0^t b_1(s)\,ds$, the solution is
\begin{eqnarray}
v_1(t)&=&B_1(t)\,,
\quad v_2(t)=\int_0^t b_2(s)\cos B_1(s)\,ds\,, \nonumber\\
 v_3(t)&=&\int_0^t b_2(s)\sin B_1(s)\,ds\,. \nonumber      
\end{eqnarray}

The action of $SE(2)$ on $\R^2\times S^1$ such that $X_1,\,X_2,\,X_3$ are
the associated fundamental vector fields turns out to be
$\Phi((\theta,\,a,\,b),(x_1,\,x_2,\,x_3))=(\bar x_1,\bar x_2,\bar x_3)$ with 
$$
\bar x_1=x_1-b \cos x_3-a \sin x_3\,,\quad \bar x_2=x_2+b \sin x_3-a \cos x_3\,,
\quad \bar x_3=x_3-\theta\,,
$$
where $(\theta,\,a,\,b)$ are the second kind canonical coordinates 
determined by the factorization $g=\exp(\theta a_1)\exp(a a_2)\exp(b a_3)$. 
The composition law is
$$(\theta,a,b)(\theta^\prime,a^\prime,b^\prime)
\!=\!(\theta+\theta^\prime,a^\prime+a \cos\theta^\prime+b \sin\theta^\prime,
b^\prime-a \sin\theta^\prime+b \cos\theta^\prime)\,.$$
Then, the general solution of (\ref{cs_unicy}) is
\begin{eqnarray}
&&\!\!\Phi((-v_1,-v_2,-v_3),(x_{10},x_{20},x_{30}))\!\!
=\!\!(x_{10}+v_3 \cos x_{30}+v_2 \sin x_{30},\cr
&&\quad\qquad\qquad\qquad\qquad
x_{20}-v_3 \sin x_{30}+v_2 \cos x_{30},x_{30}+v_1)\,,    \nonumber
\end{eqnarray}
where $v_1=v_1(t)$, $v_2=v_2(t)$ and $v_3=v_3(t)$ are given above. 

In an alternative way, as the vector fields $X_2$ and $X_3$ commute,  
there exist coordinates $(y_1,y_2,y_3)$ such 
that $X_2=\partial/\partial y_2$ and $X_3=\partial/\partial y_3$.
For instance,
$$
y_2=x_1\,\sin x_3+x_2\,\cos x_3\,,\quad y_3=x_1\,\cos x_3-x_2\,\sin x_3\,,
$$
which can be completed with $y_1=x_3$. Then, 
$$
X_1=\pd{}{y_1}+y_3\pd{}{y_2}-y_2\pd{}{y_3}\,.
$$
The control system of interest, whose solutions are again the integral curves 
of the time-dependent vector field $b_1(t)\, X_1+b_2(t)\, X_2$, reads 
\begin{equation}
\dot y_1=b_1(t)\,,\quad \dot y_2=b_1(t) y_3+b_2(t)\,,\quad \dot y_3=-b_1(t)y_2\,.
\label{cs_unicy_cor_y}
\end{equation}
Now, the expression of the previous action in terms of 
the coordinates $(y_1,\,y_2,\,y_3)$ is
\begin{eqnarray}
\Phi((\theta,\,a,\,b),\,(y_1,\,y_2,\,y_3))
&=&(y_1-\theta,\,y_2 \cos\theta-y_3\sin\theta-a\cos\theta+b\sin\theta,\,  \nonumber\\
&&\qquad\quad y_2\sin\theta+y_3 \cos\theta-a\sin\theta-b\cos\theta)\,, \nonumber
\end{eqnarray}
and hence, the general solution of (\ref{cs_unicy_cor_y}) is 
$\Phi((-v_1,-v_2,-v_3),(y_{10},y_{20},y_{30}))$, i.e.,
\begin{eqnarray}
&&y_1=y_{10}+v_1\,,\nonumber\cr
&&y_2=y_{20}\cos v_1+y_{30}\sin v_1
+v_2\cos v_1+v_3\sin v_1\,,\nonumber\cr
&&y_3=y_{30}\cos v_1-y_{20}\sin v_1
+v_3\cos v_1-v_2\sin v_1\,,\nonumber
\ea
where the $v_i$'s are those given above. 

\subsection*{{\bf 6.2.} Brockett nonholonomic control system.}

Another interesting example introduced by Brockett,
when dealing with problems of optimal control and its relation 
with singular Riemannian geometry, is related with 
the three-dimensional Heisenberg--Weyl group $H(3)$, which is the 
lowest-dimensional non-Abelian nilpotent Lie group.
Such a system is very often considered as one of the
prototypical examples relating control theory and 
extremal problems in sub-Riemannian geometry, 

It is the control system in $\R^3$, with coordinates $(x,\,y,\,z)$
\begin{equation}
\dot x=b_1(t)\,,\quad\dot y=b_2(t)\,,\quad\dot z=b_2(t) x-b_1(t) y\,,
\label{Heis_Brock_Dai}
\end{equation}
where the functions $b_1(t)$ and $b_2(t)$ are regarded as the controls. 
The solutions of this system are the integral curves 
of the time-dependent vector field $b_1(t)\, X_1+b_2(t)\, X_2$, 
with 
$$
X_1=\pd{}{x}-y\pd{}{z}\,,\quad\quad X_2=\pd{}{y}+x\pd{}{z}\,.
$$
The Lie bracket $$X_3=[X_1,\,X_2]=2 \pd{}{z}$$ is linearly independent from $X_1,\,X_2$, 
and the set $\{X_1,\,X_2,\,X_3\}$ close on the Lie algebra defined by 
\begin{equation}
[X_1,\,X_2]=X_3\,,\quad\quad [X_1,\,X_3]=0\,,\quad\quad[X_2,\,X_3]=0\,,
\label{comm_Heis_camp_vec}
\end{equation}
isomorphic to the Lie algebra $\goth h(3)$ of the Heisenberg--Weyl group $H(3)$.

The Lie algebra $\goth h(3)$ has a basis $\{a_1,\,a_2,\,a_3\}$ for which
the Lie products are
$$
[a_1,\,a_2]=a_3\,,\quad [a_1,\,a_3]=0\,,\quad [a_2,\,a_3]=0\,.
$$
A generic Lie system for the particular case of $H(3)$ takes the form 
$$
R_{g(t)^{-1}*g(t)}(\dot g(t))=-b_1(t)a_1-b_2(t)a_2-b_3(t)a_3\,,
$$
and we are now interested in the one with $b_3(t)=0$ for all $t$, i.e., 
\begin{equation}
R_{g(t)^{-1}*g(t)}(\dot g(t))=-b_1(t)a_1-b_2(t)a_2\,.
\label{eq_grup_H3_b3nulo}
\end{equation}
Writing the solution starting from the identity of (\ref{eq_grup_H3_b3nulo})
as the product of exponentials
$$
g(t)=\exp(-v_1(t)a_1)\exp(-v_2(t)a_2)\exp(-v_3(t)a_3)\ ,
$$
and applying the Wei-Norman formula (\ref{eq_Wei_Nor}) 
we find the system of differential equations 
$$
\dot v_1=b_1\,,\quad\dot v_2=b_2\,,\quad\dot v_3=b_2\,v_1\,,
$$
with initial conditions $v_1(0)=v_2(0)=v_3(0)=0$. 
The solution can be found immediately:
\begin{eqnarray}
v_1(t)&=&\int_0^t b_1(s)\,ds\,,
\quad v_2(t)=\int_0^t b_2(s)\,ds\,,\nonumber\\
 v_3(t)&=&\int_0^t b_2(s)\int_0^s b_1(r)\,dr\,ds\,.
\label{sol_vs_Heis}
\end{eqnarray} 

The preceding solution can be used in order to find the general solution 
of the given system (\ref{Heis_Brock_Dai}). We only need to 
find a suitable parametrization of the Lie group $H(3)$, 
and the expression of the group action with respect to 
which the original vector fields are the fundamental vector fields. 
If we take the canonical coordinates of second class 
defined by $g=\exp(a a_1)\exp(b a_2)\exp(c a_3)$, when $g\in H(3)$,
it can be shown that such an action reads
\begin{eqnarray*}
\Phi:H(3)\times\R^3&\longrightarrow& \R^3                       \nonumber\\
((a,\,b,\,c),\,(x,\,y,\,z))&\longmapsto&(x-a,\,y-b,\,z+a y-b x-a b-2 c)\,,
\end{eqnarray*}
 the group law being expressed as
$$
(a,\,b,\,c)(a^\prime,\,b^\prime,\,c^\prime)
=(a+a^\prime,\,b+b^\prime,\,c+c^\prime-b a^\prime)\,.
$$
Then, the general solution of (\ref{Heis_Brock_Dai}) is 
\begin{eqnarray}
\Phi((-v_1,\,-v_2,\,-v_3),(x_0,\,y_0,\,z_0))
&=&(x_0+v_1,\,y_0+v_2,\,z_0+x_0 v_2\nonumber\\&&\qquad\quad-y_0 v_1-v_1 v_2+2 v_3)\,,\nonumber
\end{eqnarray}
where $v_1=v_1(t)$, $v_2=v_2(t)$, and $v_3=v_3(t)$ are given 
by (\ref{sol_vs_Heis}).

\subsection*{{\bf 6.3.} Hopping robot in flight phase.}

Next we consider another example coming from a physical model:
a hopping robot in flight phase, which has been 
studied, e.g., in \cite{MurSas93}. The system consists of
a body with an actuated leg that can rotate and extend. 
The coordinates are $(\psi,\,l,\,\theta)$, describing  the 
body angle, leg extension and leg angle of the robot, respectively. 
The constant $m_l$ is the mass of the leg, and the mass of the body 
is taken to be one. The interest is focused on the behaviour 
of the system for small elongation, that is, near $l=0$. 
Precisely, the approximation of the system in the neighbourhood 
of $l=0$ will lead to a Lie system related to the Heisenberg--Weyl group, 
and therefore related also to the previous example. 

The controls of the system are the leg angle and extension velocities. 
The control system takes the form \cite{MurSas93}
\begin{equation}
\dot \psi=b_1(t)\,,\quad \dot l=b_2(t)\,,
\quad \dot\theta=-\frac{m_l (l+1)^2}{1+m_l (l+1)^2} b_1(t)\,,
\label{cont_syst_hop_rob}
\end{equation}
whose solutions are the integral curves of the time-dependent vector field
$b_1(t)\, Y_1+b_2(t)\, Y_2$, where now
$$
Y_1=\pd{}{\psi}-\frac{m_l (l+1)^2}{1+m_l (l+1)^2}\pd{}{\theta}
\,,\quad\quad Y_2=\pd{}{l}\,.
$$
However, the system (\ref{cont_syst_hop_rob}) cannot be considered as 
a Lie system, since the iterated Lie brackets 
$$
\left[Y_2,\,\left[Y_2,\,\dots \left[Y_2,\,Y_1\right] \cdots \right]\right]
$$
generate at each step vector fields linearly independent from 
those obtained at the previous stage. Notwithstanding, in order
to steer the original system by sinusoids, it was  
proposed in \cite{MurSas93} to take the Taylor 
approximation, linear in $l$, of the system, that is,
\begin{equation}
\dot \psi=b_1(t)\,,\quad \dot l=b_2(t)\,,\quad \dot\theta=-(k_1+k_2 l) b_1(t)\,,
\label{sist_Hop_rob_approx}
\end{equation}
where the constants $k_1$ and  $k_2$ are defined as
$$
k_1=\frac{m_l}{1+m_l}\,,\quad k_2=\frac{2 m_l}{(1+m_l)^2}\,,
$$
and then the vector fields become
\begin{equation}
X_1=\pd{}{\psi}-(k_1+k_2 l)\pd{}{\theta}\,,\quad\quad X_2=\pd{}{l}\,.
\label{vf_hop_rob_lin_appr}
\end{equation}
Now, the new vector field
$$
X_3=[X_1,\,X_2]=k_2\pd{}{\theta}
$$
closes, jointly with $X_1$, $X_2$, the Lie algebra (\ref{comm_Heis_camp_vec}),
so that (\ref{sist_Hop_rob_approx}) can be regarded as a Lie system 
with associated Lie algebra $\goth h(3)$. 

If we use the previously defined canonical coordinates $(a,\,b,\,c)$ of 
second kind for parametrizing the group $g\in H(3)$, the corresponding (local) 
action to our Lie system reads
\begin{eqnarray*}
\Phi:H(3)\times M&\longrightarrow& M                    \nonumber\\
((a,\,b,\,c),\,(\psi,\,l,\,\theta))&\longmapsto&(\psi-a,\,l-b,\,\theta+k_2(a l-c-a b)+a k_1)\,,
\end{eqnarray*}
where $M$ is a suitable open set of $\R^3$.
Then, the general solution of the system (\ref{sist_Hop_rob_approx}) 
can be written, for $t$ small enough, as
$$
\Phi((-v_1,\,-v_2,\,-v_3),(\psi_0,\,l_0,\,\theta_0))
=(\psi_0+v_1,\,l_0+v_2,\,\theta_0+k_2(v_3-v_1 l_0-v_1 v_2)-k_1 v_1)\,,
$$
where $v_1=v_1(t)$, $v_2=v_2(t)$, and $v_3=v_3(t)$ are given 
by (\ref{sol_vs_Heis}). This result can be checked by
direct integration.

\subsection*{{\bf 6.4.} Kinematics of the generalization 
of the elastic problem of Euler as a Lie system.}

Recently, Jurdjevic has generalized the so-called elastic problem of Euler
to homogeneous spaces of constant curvature embedded in a three dimensional 
Euclidean space, in order to study certain integrable Hamiltonian 
systems from the point of view of optimal control theory \cite{Jur95}. 
We will only deal with the kinematic equations of such systems, 
wich turn out to be Lie systems. 

The system of interest is the control system with configuration space $\R^3$, 
and coordinates $(x_1,\,x_2,\,x_3)$, given by
\begin{eqnarray}
 \dot x_1&=&-b_1(t)x_2-b_2(t)x_3\,,\nonumber\\
 \dot x_2&=&b_1(t) x_1+b_3(t) x_3\,,\label{syst_jurd}\\
 \dot x_3&=&\epsilon\,(b_2(t) x_1-b_3(t) x_2)\,,\nonumber
\end{eqnarray}
where $\epsilon=\pm 1,0$. Its solutions are the integral curves of the 
time-dependent vector field 
$b_1(t)\, X_1(x)+b_2(t)\, X_2(x)+b_3(t)\, X_3(x)$, where
\begin{eqnarray}
 X_1&=&x_1\,\pd{}{x_2}-x_2\,\pd{}{x_1}\,,\nonumber\\
 X_2&=&\epsilon\,x_1\,\pd{}{x_3}-x_3\,\pd{}{x_1}\,,\label{input_vf_Ge}\\
 X_3&=&x_3\,\pd{}{x_2}-\epsilon\,x_2\,\pd{}{x_3}\,.\nonumber
\end{eqnarray}
These vector fields satisfy the commutation relations
$$
[X_1,\,X_2]=X_3\,,\qquad [X_2,\,X_3]=\epsilon\,X_1\,,\qquad [X_3,\,X_1]=X_2\,,
$$ 
and hence they generate a Lie algebra isomorphic to $\goth g_\epsilon$ of the
Lie group $G_\epsilon$, given by $G_0=SE(2)$, $G_1=SO(3)$ and $G_{-1}=SO(2,1)$. 
Therefore, the case $\epsilon=0$ essentially reduces to the first example studied
in this section.
We take a  basis $\{a_1,\,a_2,\,a_3\}$ of $\goth g_\epsilon$ in which
the Lie products read 
$$
[a_1,\,a_2]=a_3\,,\qquad [a_2,\,a_3]=\epsilon\,a_1\,,\quad [a_3,\,a_1]=a_2\,,
$$ 
and define the signature-dependent trigonometric functions 
(see, e.g., \cite{BallHerOlmSan93}):   
$$
\quad\quad C_{\epsilon}(x)=\left\{\begin{array}{ll}
        \cos {x} &\quad \epsilon=1 \cr
        1        &\quad \epsilon=0 \cr
        \cosh{x} &\quad \epsilon=-1 
\end{array}\right. 
\quad
S_{\epsilon}(x)=\left\{\begin{array}{ll}
        \sin {x} &\quad \epsilon=1 \cr
        x        &\quad \epsilon=0 \cr
        \sinh{x} &\quad \epsilon=-1 
\end{array}\right. 
\quad
\nonumber
$$
and $$
T_{\epsilon}(x)=\frac{S_{\epsilon}(x)}{C_{\epsilon}(x)}\,,
$$
which satisfy
\begin{eqnarray}
C_\epsilon(x+y)&=&C_\epsilon(x)C_\epsilon(y)-\epsilon\,S_\epsilon(x)S_\epsilon(y)\,,\nonumber\\
        S_\epsilon(x+y)&=&C_\epsilon(x)S_\epsilon(y)+S_\epsilon(x)C_\epsilon(y)\,, 
                                                                                \nonumber
\end{eqnarray}
as well as $C_\epsilon^2(x)+\epsilon\,S_\epsilon^2(x)=1$,
and
\begin{eqnarray}
\frac{d C_\epsilon(x)}{dx}&=&-\epsilon\, S_\epsilon(x)\,,
\qquad\frac{d S_\epsilon(x)}{dx}=C_\epsilon(x)\,,  \nonumber \\                     
 \frac{d T_\epsilon(x)}{dx}&=&1+\epsilon\, T^2_\epsilon(x)
=\frac 1{C^2_\epsilon(x)}\,. \nonumber
\end{eqnarray}
Writing the solution of the problem associated to (\ref{syst_jurd}) 
in the group $G_\epsilon$ as the 
product $g(t)=\exp(-v_1(t)a_1)\exp(-v_2(t)a_2)\exp(-v_3(t)a_3)$, and 
using the Wei--Norman formula (\ref{eq_Wei_Nor}), 
we obtain the system of differential equations for $v_1(t),\,v_2(t)$ and $v_3(t)$:
\ba
&&\dot v_1=b_1+\epsilon\,T_\epsilon(v_2)(b_3\,\cos v_1+b_2\,\sin v_1)\,,\nonumber\\
&&\dot v_2=b_2\,\cos v_1-b_3\,\sin v_1\,,\nonumber\\
&&\dot v_3=\frac{b_3\,\cos v_1+b_2\,\sin v_1}{C_\epsilon(v_2)}\,,\nonumber
\ea
with $v_1(0)=v_2(0)=v_3(0)=0$. 
Other possible reorderings of the factorization in exponentials 
will give rise to similar systems of equations.
For $\epsilon=\pm 1$ the group $G_\epsilon$ is simple and none of 
the Wei--Norman systems can be integrated by quadratures in a general case.
For the particular case treated by  Jurdjevic \cite{JurSus72} we must put  
(with our notation) $b_1(t)=1$, $b_2(t)=0$ and $b_3(t)=k(t)$. 

\subsection*{{\bf 6.5.} Reduction of Lie systems in control theory.}

Finally, we would like to point out that there exist a technique 
for reducing the problem of solving a given Lie system 
in a Lie group $G$ to solving a similar Lie system but 
in a subgroup $H$, provided a particular solution of the problem 
corresponding to the former in an associated homogeneous space $G/H$ is known
(see, e.g., \cite{CarRamGra}).
This reduction procedure can be shown to be useful as well in the study of 
the particular kind of drift-free control systems, linear in the control functions, 
which in addition are Lie systems.

Take for instance, in the simplified model of maneuvering 
an automobile discussed before, the subgroup
$H=\{(0,\,0,\,b)\}$. In these coordinates,  
$\tau\!:\! SE(2)\!\rightarrow\! SE(2)/H$ is 
$\tau(\theta,\,a,\,b)=(\theta,\,a)$. 
Taking coordinates $(z_1,\,z_2)$ in $M=SE(2)/H$, we have that the 
left action of $SE(2)$ on $M$ is given by
$$
\Phi((\theta,\,a,\,b),(z_1,\,z_2))=(z_1+\theta,\,z_2+a \cos z_1+b \sin z_1)\ .
$$
The fundamental vector fields with respect to this action are
$$
X_1^H=-\pd{}{z_1}\,,\quad
X_2^H=-\cos z_1\pd{}{z_2}\,,\quad
X_3^H=-\sin z_1\pd{}{z_2}\,,
$$
which satisfy $$[X_1^H,\,X_2^H]=X_3^H\,, \quad [X_2^H,\,X_3^H]=0\,,\quad  
[X_1^H,\,X_3^H]=-X_2^H\ ,$$ and the equations on the homogeneous 
space $M$ to be solved are 
$$
\dot z_1=-b_1(t)\,,\quad \dot z_2=-b_2(t) \cos z_1\,.
$$ 
Assume we have a curve on $SE(2)$, $g_1(t)$, such that the coordinates
of its projection $\tau(g_1(t))=(z_1(t),\,z_2(t))$ satisfy the previous 
equations. For example, we take $g_1(t)=(z_1(t),\,z_2(t),\,0)$. 
Then, we can reduce the problem to solving an equation on $H$,
which takes the form
$$
\dot b(t)=b_2(t) \sin z_1(t)\,,
$$
which is just a Lie system for the additive group of the real line.

As a second and last example, consider again the kinematic equations
of the generalized elastic problem of Euler treated previously. 
In this case, however, it is advantageous to consider instead of the
Lie group $G_\epsilon$, its universal covering $\bar G_\epsilon$, 
in order to perform the reduction. We have that $\bar G_1=SU(2)$, 
$\bar G_{-1}=SU(1,1)$, and $\bar G_0=SE(2)$. 
The elements of the group $\bar G_\epsilon$ can be parametrized 
by four real numbers $(a,\,b,\,c,\,d)$ such 
that $a^2+b^2+\epsilon(c^2+d^2)=1$,  the group law
$$
(a,\,b,\,c,\,d)(a^\prime,\,b^\prime,\,c^\prime,\,d^\prime)=
(a^{\prime\prime},\,b^{\prime\prime},\,c^{\prime\prime},\,d^{\prime\prime})
$$
being given by
\begin{eqnarray}
&& a^{\prime\prime}=a a^\prime-b b^\prime-\epsilon\,(c c^\prime+d d^\prime)\,,
\quad b^{\prime\prime}=b a^\prime+a b^\prime-\epsilon\,(d c^\prime-c d^\prime)\,,\nonumber\\
&& c^{\prime\prime}=c a^\prime+d b^\prime+a c^\prime-b d^\prime\,,      
\quad d^{\prime\prime}=d a^\prime-c b^\prime+b c^\prime+a d^\prime\,,   \nonumber
\end{eqnarray}
see also \cite{Bro72,Stu64}. To perform the reduction we take the subgroup $H$
generated by $a_1$. The projection   
$\tau\!:\! \bar G_\epsilon\!\rightarrow\! \bar G_\epsilon/H$ is defined by
$$
\tau(a,\,b,\,c,\,d)=\left(\frac{a c-b d}{a^2+b^2},\,\frac{b c+a d}{a^2+b^2}\right)\,.
$$ 
Taking coordinates $(z_1,\,z_2)$ in $M=\bar G_\epsilon/H$, the left action 
of $\bar G_\epsilon$ on $M$ reads 
$$
\Phi((a,\,b,\,c,\,d),(z_1,\,z_2))=\left(\frac{N_1}{D},\,\frac{N_2}{D}\right)\,,
$$
where
\begin{eqnarray}
N_1\!\!\!\!&=&\!\!\!\!(a^2-b^2-\epsilon\,(c^2-d^2))z_1-2(a b+\epsilon\, c d)z_2
+(a c-b d)(1-\epsilon\,(z_1^2+z_2^2))\,,                                        \nonumber\\
 N_2\!\!\!\!&=&\!\!\!\!2(a b-\epsilon\, c d)z_1+(a^2-b^2+\epsilon\,(c^2-d^2))z_2
+(a d+b c)(1-\epsilon\,(z_1^2+z_2^2))\,,                                        \nonumber\\
 D\!\!\!\!&=&\!\!\!\!a^2+b^2-2 \epsilon\,((a c+b d)z_1+(a d-b c)z_2)
+\epsilon^2(c^2+d^2)(z_1^2+z_2^2)\,.                                            \nonumber
\end{eqnarray}
The fundamental vector fields with respect to this action are
\begin{eqnarray}
 X_1^H&=&z_2\pd{}{z_1}-z_1\pd{}{z_2}\,,\nonumber\\
 X_2^H&=&-\frac 1 2(1+\epsilon\,(z_1^2-z_2^2))\pd{}{z_1}-\epsilon\,z_1 z_2\pd{}{z_2}\,,                                                                                \nonumber\\
X_3^H&=&-\epsilon\,z_1 z_2\pd{}{z_1}-\frac 1 2(1-\epsilon\,(z_1^2-z_2^2))\pd{}{z_2}\,,                                                                               \nonumber
\end{eqnarray}
which satisfy 
$$
[X_1^H,\,X_2^H]=X_3^H\,, \quad [X_2^H,\,X_3^H]=\epsilon\,X_1^H\,,\quad [X_1^H,\,X_3^H]=-X_2^H\ ,
$$ 
and the equations on the homogeneous space $M$ to be solved are 
\begin{eqnarray}
&& \dot z_1=b_1(t)z_2-\frac 1 2 b_2(t)(1+\epsilon\,(z_1^2-z_2^2))-b_3(t)\,\epsilon\,z_1 z_2\,,
                                                                                \nonumber\\ 
&& \dot z_2=-b_1(t)z_1-b_2(t)\,\epsilon\,z_1 z_2-\frac 1 2 b_3(t)(1-\epsilon\,(z_1^2-z_2^2))\,.
                                                                                \nonumber
\end{eqnarray}
Assume we have a curve $g_1(t)$ on $\bar G_\epsilon$, such that 
its projection $\tau(g_1(t))=(z_1(t),\,z_2(t))$ satisfy the previous 
equations. We can take, for example, 
$$
g_1(t)=\frac{(1,\,0,\,z_1(t),\,z_2(t))}{\sqrt{1+\epsilon\,(z^2_1(t)+z^2_2(t))}}\,.
$$
Then, we can reduce the problem to an equation on $H$: if its solution is of the form
$h(t)=(\cos(v(t)/2),\,\sin(v(t)/2),\,0,\,0)$, then $v(t)$ satisfies
$$
\dot v(t)=-b_1(t)+\epsilon\,(b_3(t) z_1(t)-b_2(t) z_2(t))\,.
$$

\section*{Acknowledgements}
First author has been partially supported by the Spanish DGI, 
project BFM2000-1066-C03-01. 
Second author supported by a FPI grant of the Spanish MCyT.

\end{document}